\documentclass[letterpaper,twocolumn,10pt]{article}
\usepackage{usenix}

\usepackage{url}
\usepackage{verbatim}
\usepackage[utf8]{inputenc}
\usepackage{microtype}
\usepackage{booktabs}
\usepackage{amssymb}
\usepackage{alltt}
\usepackage{upquote}
\usepackage{fancyvrb}
\usepackage{graphicx}
\usepackage{subfigure}
\usepackage{multirow}

\begin{document}

\date{}

\title{\Large \bf Automated Password Extraction Attack on Modern Password Managers}

\author{
{\rm Raul Gonzalez}\\
Carnegie Mellon University
\and
{\rm Eric Y. Chen}\\
Carnegie Mellon University
\and
{\rm Collin Jackson}\\
Carnegie Mellon University
} 

\maketitle

\thispagestyle{empty}

\subsection*{Abstract}

To encourage users to use stronger and more secure passwords, modern web browsers offer users password management services, allowing users to save previously entered passwords locally onto their hard drives. We present Lupin, a tool that automatically extracts these saved passwords without the user's knowledge. Lupin allows a network adversary to obtain passwords as long as the login form appears on a non-HTTPS page. Unlike existing password sniffing tools, Lupin can obtain passwords for websites users are not visiting. Furthermore, Lupin can extract passwords embedded in login forms with a destination address served in HTTPS. To determine the number of websites vulnerable to our attack, we crawled the top 45,000 most popular websites from Alexa's top website list and discovered that at least 28\% of these sites are vulnerable. To further demonstrate the feasibility of our attack, we tested Lupin under controlled conditions using one of the authors' computers. Lupin was able to extract passwords from 1,000 websites in less than 35 seconds. We suggest techniques for web developers to protect their web applications from attack, and we propose alternative designs for a secure password manager.

\section{Introduction}

Users' passwords have often been the weakest link in securing modern 
web applications. Even when a web application is reinforced with the 
most sophisticated security features, an adversary can often 
compromise users' accounts by launching a brute force attack on their 
login passwords. A study done by Florencio et al. in 2007 showed that a vast majority of 
web passwords consists solely of lower-case alphabetical 
characters~\cite{www07_florencio}. The habit of using easy-to-remember 
passwords nullifies any defenses put in place by web developers and 
greatly increases the risk of user accounts being compromised. 

In order to encourage users to choose unique, secure passwords for 
websites without burdening them with remembering each password, 
browser vendors have augmented browsers with a service called 
\emph{password manager}. When a user first enters her password to a 
website, the password manager will prompt her for permission to 
save the password locally. If the 
permission is granted, the browser will store this password and 
refill it when the user revisits the same web page. Over the past 
decade, password managers have gone through various security 
analysis~\cite{passhash,passpet}, and browser vendors have taken several 
steps to ensure that the password manager cannot be abused by the 
attacker. However, despite all of the existing defenses,
we describe a new attack that allows 
an adversary with network capability to automatically extract 
passwords stored in the user's password manager. 

To demonstrate our attack, we created a tool called Lupin that allows an attacker 
connected to a wireless network to steal the saved passwords of other 
users within the same network. Lupin operates in four steps; first, 
Lupin establishes itself as the victim's network gateway by launching 
an ARP spoofing attack. Second, Lupin waits for the victim to
request to any unencrypted web page, then piggybacks the attack code 
onto the response. The attack code consists of a large number of 
iframes, each pointing to a different website that the adversary wants 
to extract passwords from. Third, Lupin waits for the victim's browser 
requests for these framed pages and responds to each request with a 
bogus page containing a login form and a piece of malicious JavaScript 
code. Finally, when the victim's password manager fills in the 
passwords into the bogus login forms, the malicious JavaScript code 
will extract the information and send it back to the attacker. 
Lupin is superior to a conventional network eavesdropper, because
Lupin can obtain passwords submitted to an HTTPS web page. 
Since it is a common practice 
for websites to serve public content in HTTP and redirect users to 
HTTPS pages when they decide to log in, Lupin can gather passwords 
associated with these websites, while a passive eavesdropper cannot. 

To determine the number of websites vulnerable to Lupin, we crawled 
45,000 most popular websites according to Alexa's top website list. 
We discovered that at least 28\% of all sites are vulnerable to Lupin. 
Additionally, we measured the performance of Lupin under controlled 
conditions using one of the authors' computers. Lupin was able to 
extract passwords from 1,000 websites in less than 35 seconds. To 
protect users from Lupin, we propose a fix for Chrome and Firefox's 
current password managers that maximizes usability while protecting 
users' HTTPS passwords from being stolen by the attacker.

The remainder of this paper is organized as follows. 
Section~\ref{sec:background} briefly describes the background of our 
attack. Section~\ref{sec:attack} details our attack. 
Section~\ref{sec:evaluation} evaluates the feasibility and the impact 
of our attack. We propose possible defenses in 
Section~\ref{sec:defense}. Section~\ref{sec:related} describes related work,
and lastly, Section~\ref{sec:conclusion} concludes.

\section{Background}
\label{sec:background}
The browser's password manager offers an intuitive way for users to 
store unique and secure passwords for each website they visit. 
However, by shifting the responsibility of identifying the appropriate 
login forms away from users, password managers become an attraction 
for online miscreants. To protect password managers from malicious 
online entities, browser vendors made an effort to ensure that users' 
passwords are not exposed to attackers. Ideally, browsers must only 
present users' login credentials to legitimate login forms. However, 
different browsers have different notions of when to auto-fill a 
password. Table~\ref{tab:browser_survey} describes how different 
browsers decide the appropriate location to auto-fill passwords. When 
a web page presents the user with a login form, the browser generally 
considers three factors before deciding to auto-fill the form with the 
user's login credentials. We describe these three factors in detail 
below.

\begin{itemize}
	\item \textbf{URL} -- Intuitively, the most important factor 
in deciding whether to auto-fill a login form is the location of the 
web page containing the login form. When the user enters her password 
for the first time, the browser will record the location of the web 
page embedding the login form; we call this the \emph{source 
location}. The next time the user visits a web page containing a login 
form, the browser will compare its location with the source locations 
of existing login credentials in the database. If the two locations 
match to a certain degree, the browser will proceed to the next step. 
Most of the browsers (with exception to IE) match the source locations 
of the login forms based on their origins, while IE matches their 
paths. The security argument accompanying origin-based matching is 
that path-based matching does not add any additional security benefits 
against same-origin attackers, since same-origin attackers already have 
full JavaScript execution capabilities~\cite{beware}. 
	
	\item \textbf{User action} -- Two of the five browsers we 
studied, namely IE and Opera, require users to manually initiate the 
password manager. For IE, the user must enter the first character of 
her username in order to trigger the auto-fill process. Similarly, 
Opera requires the user to manually press the auto-fill button or 
enter a special character sequence (ctrl + enter) to begin the auto-fill 
process. We discuss in Section~\ref{sec:attack} how these 
behaviors affect our attack.
	
	\item \textbf{DOM} -- In addition to URL and user action 
requirements, many browsers impose additional requirements on the DOM 
to ensure that the password is not exposed to the adversary. One of 
the most common requirements is that the destination address of the 
login form (i.e., the target of the form post) must coincide with the 
destination address of the initial login form where the password was 
stored; unfortunately, this defense was recently shown to be 
ineffective~\cite{selfexfiltration}. Besides checking for the 
destination address of the form post, Safari has a unique requirement 
that does not allow passwords to be auto-filled into iframes. 
\end{itemize}

\begin{table*}
\centering
\begin{tabular}{|c|p{4cm}|p{4cm}|p{4cm}|}
\hline
\textbf{Browsers} & \textbf{URL requirement} & \textbf{User action 
requirement}& \textbf{DOM requirement}\\
\hline
Internet Explorer & Source address's origin and path must match. & Must 
enter the first character of the username & None \\
\hline
Opera & Source address's origins must match. & Must click on the ``auto-fill 
button'' or press ``Control + Enter''  &  Destination address's 
origins must match. The ``name'' attribute of the input fields must 
match.\\
\hline
Safari & Source address's origins must match.  & None & Login form must 
be inside the top-level frame.\\
\hline
Firefox &  Source address's origins must match.  & None &  Destination 
address's origins must match. \\
\hline
Chrome & Source address's origins must match.  & None &  Destination 
address's origins must match. \\
\hline
\end{tabular}
\caption{Requirements for auto-filling passwords in different 
browsers where source address represents the URL of the page that 
embedded the login form and destination address represents the 
location the login form is submitted to. }
\label{tab:browser_survey}
\end{table*}

\subsection{Threat Model}
\label{sec:threat}
We proceed to describe the capability of the adversary as well as user 
behaviors assumed for the rest of this paper. We consider a standard 
\emph{network attacker}, where the adversary has the ability to 
intercept, eavesdrop, and modify any unencrypted network packets. 
However, the attacker does not have the ability to break existing 
encryption schemes in order to gain access to SSL traffic. 

We treat the user as a \emph{security paranoid} individual. That is, 
the user can distinguish HTTPS web pages from their HTTP counterparts. 
Furthermore, the user heeds all security warnings and refrains 
from logging into any HTTP pages while using an insecure network. 
However, the user may still visit other HTTP pages while using an 
insecure network without logging in.

\section{Attack}
\label{sec:attack}
In this section, we describe our attack in detail. Our attack exploits 
the weakness in the Firefox and Chrome password managers; it allows 
the network adversary to automatically explore web passwords stored in 
the victim's browser. To demonstrate the effectiveness of the attack, 
we created Lupin -- a network level, fully automated tool for password 
theft. 

\begin{figure}[ht]
\begin{center}
	
\includegraphics[width=0.9\linewidth]{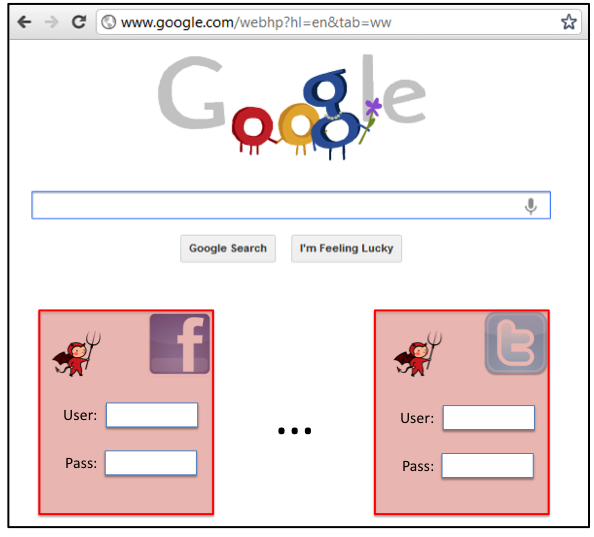}
\end{center}
\caption{Attacker controlled iframes are piggybacked onto the response 
of a benign HTTP request. These iframes are used to trigger the 
victim's password manager.}
\label{fig:frame}
\end{figure}

We provide a detailed description of our attack below. We assume the 
adversary to be a network attacker described previously in 
Section~\ref{sec:threat}. Furthermore, we assume the victim visits an 
arbitrary HTTP web page while using the insecure network. 
\begin{enumerate}
	\item The adversary waits for the victim to make a request to 
an unencrypted page served in HTTP then piggybacks onto the response 
a large number of iframes, each pointing to a different web page that 
the attacker wishes to extract passwords from, as depicted in 
Figure~\ref{fig:frame}. Web pages embedded in these iframes must be 
served in HTTP.
	\item After the victim's browser receives the tampered 
response, it will subsequently make requests for the web pages 
associated with each of the iframes. 
	\item The adversary again intercepts these requests and 
responds to each web request with a bogus web page containing a login 
form and a piece of JavaScript code. 
	\item When these bogus web pages are delivered to the victim's 
browser, they will in turn trigger the browser's password manager to 
auto-fill passwords for each of these web pages. After these login 
forms are auto-filled, the malicious JavaScript code will read the 
login credentials on the login form and send them back to the 
attacker.
\end{enumerate}    

The success of the attack rests upon our ability to deceive the 
victim's password manager into filling the user's login credentials 
into a web page that has been tampered by the adversary. However, not 
all password managers are vulnerable to this attack. Recall from 
Table~\ref{tab:browser_survey}, only Chrome and Firefox automatically 
fill in saved passwords for non-top-level frames. Since Chrome and 
Firefox currently consist of around 40\% of the browser market 
share~\cite{marketshare}, our attack poses a significant risk to a 
large portion of users. We describe below the reasons why our attack 
fails to work for other browsers.

\begin{itemize}	
	\item \textbf{IE and Opera} -- Both IE and Opera require user 
interaction before auto-filling any login credentials. However, due to 
the fully automated nature of our attack, we cannot generate or forge 
the user interactions required for either IE or Opera.
	
	\item \textbf{Safari} -- Although Safari's password manager 
does not require any user interaction before auto-filling login 
credentials, it only auto-fills login forms located inside the top-level 
frame; that is, Safari will not auto-fill any login forms inside 
our injected iframe. One way to circumvent this is to use popup 
windows instead of iframes, but this would significantly reduce the 
stealthiness of our attack. 

\end{itemize}

\subsection{Lupin}
We implemented our attack as an automated tool called Lupin, which 
consists of 800 lines of Python and JavaScript code. To use Lupin, the 
adversary simply connects to a wireless network. Next, Lupin scans for 
all available nodes in the network, then proceeds to launch an ARP 
spoofing attack on each node to impersonate the network gateway (this 
step is done using the ``dsniff'' package in Linux). After 
establishing itself as the bogus network gateway, the adversary can 
then carry out the attack described previously in 
Section~\ref{sec:attack}. We provide a more thorough description of 
the tool below.

\subsubsection{Scalability}

\begin{figure}[ht]
\begin{center}
	
\includegraphics[width=0.9\linewidth]{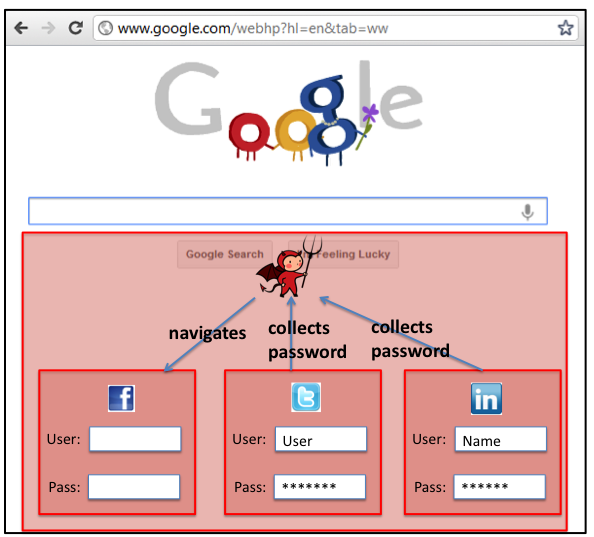}
\end{center}
\caption{Lupin arranges the injected frames in a hierarchical fashion. 
Each target web page is loaded asynchronously.}
\label{fig:scale}
\end{figure}

Although some adversaries are only interested in passwords from a 
small subset of websites, we believe the effectiveness of the attack 
would be significantly increased if the attacker was able to extract a 
large number of passwords rapidly. Recall from Figure~\ref{fig:frame} that 
in order to efficiently extract a large number of passwords, one must 
create multiple iframes and perform the attack in parallel. 
Intuitively, one could create one iframe for each target web page. 
However, this would create a burst of traffic on the network and consume 
a huge amount of memory on the victim's browser, making 
the attack easily detectable. Lupin avoids this problem by organizing 
the iframes in a hierarchical structure, as depicted in 
Figure~\ref{fig:scale}. The top-level iframe holds the bulk of the 
attack logic. It dynamically spawns child frames to trigger the 
victim's password manager, then collects the user's credentials before 
navigating the child frames to the next target web page. After 
exploring all of the target web pages, the top-level frame bundles all 
of the stolen data into a single web request and forwards it to the 
adversary.

\subsubsection{Stealth}
\label{sec:stealth}

\begin{figure*}[ht]
		\centering
		\subfigure[Chrome refresh animation.]{
                \centering
                \includegraphics[width=2.2in]{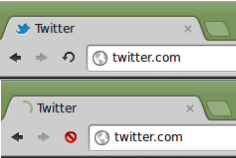}
                \label{fig:chrome_refresh}
        }%
        ~ 
etc. 
		\subfigure[Firefox refresh animation.]{
                \centering
                \includegraphics[width=2.2in]{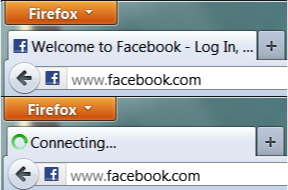}
                \label{fig:ff_refresh}
        }

        \caption[]{Refresh animation induced by running our attack 
code in a background tab.}
		\label{fig:refresh}
\end{figure*}

Lupin is designed to provide maximum stealth to the adversary. First, 
the malicious iframes are made to be hidden from the victims. This can 
be done in several ways, such as by making the iframes transparent or 
by making the size of the iframes one pixel~\cite{framebusting}. 
Second, our code detects if the user is currently focused on the 
browser tab or window containing the attack code and executes only if 
the tab or window is out of focus. Both Chrome and Firefox deploy a 
status bar that informs the user of any outgoing web requests; 
however, by running the attack in a background tab, the status bar is 
effectively hidden from the victim. Unfortunately, there is a minor 
drawback for running the attack code in the background tab; that is, 
it triggers the browser's refresh animation when it navigates the 
malicious iframes. Figure~\ref{fig:refresh} illustrates how the 
refresh animation is seen by the user. We believe this is not a major 
weakness, as many legitimate web pages (such as Gmail) already 
periodically refresh themselves. 

\section{Evaluation}
\label{sec:evaluation}
In this section, we evaluate our automated password extraction tool 
Lupin. Our evaluation is twofold; first, we measure the efficiency of 
Lupin under controlled laboratory conditions and evaluate the 
effectiveness of our attack in terms of number of websites explored 
per minute (WPM). Second, we conduct an extensive survey on the 45,000 
most popular websites from Alexa's top website list and measure 
the percentage of websites vulnerable to our attack. We summarize our 
results below.

\subsection{Performance}
To evaluate the effectiveness of Lupin in a real-world scenario, we 
tested Lupin using one of the authors' computers. The victim's browser 
and Lupin itself were located inside a virtual machine. We attempted 
to simulate a normal victim's browsing behavior by using two browser 
tabs to visit popular websites such as Facebook and Gmail 
simultaneously. We were able to explore 1,000 web pages in a period of 
35 seconds (around 2,000 WPM), with no noticeable performance 
degradation. In our study, we programmed Lupin to wait for 100 
milliseconds after a target page has finished loading, then check to 
see if the login form was auto-filled. This makes the result from our 
measurement a conservative estimate, because most browsers take less 
than 100 milliseconds to auto-fill a password. Furthermore, the speed 
could be increased if our attack code was executed inside multiple tabs 
as opposed to one. Finally, running Lupin on a host OS rather than a 
virtual machine should also improve its performance.

One interesting challenge we faced was that the user may navigate away 
from the page executing our attack code. To combat this, Lupin detects 
whether a web page is currently running in the background and only 
executes the attack code inside background tabs. Furthermore, Lupin has
the option to simulate the refresh behavior of a normal website such as Gmail;
this is achieved by issuing periodic refreshes, with each refresh lasting no
more than a few seconds. However, one downside of reducing the refresh rate
of Lupin is that the crawling speed is also decreased.

\subsection{Vulnerability Coverage}

Since Lupin cannot obtain passwords stored on HTTPS web pages, it is important to measure the ratio of websites vulnerable to our attack. To obtain this information, we created a web crawler that surveyed the 45,000 most popular websites from Alexa's top website list. In our survey, we considered a website vulnerable if it contained a login form served in HTTP and did not have the autocomplete attribute set to ``off''. Our results are described in Table~\ref{tab:crawler}. Out of the 45,000 websites we surveyed, atleast 28\% of them were vulnerable to Lupin. Some examples of vulnerable websites include Facebook, Twitter, LinkedIn, and GoDaddy. 

Some of the websites we surveyed used JavaScript to dynamically create links, forms, and other HTML content. Parsing and analyzing these pages using a basic web crawler was difficult. Therefore, to avoid false positives, we marked these websites (11,584 in total) as not vulnerable. Furthermore, we discovered that at least 12\% of the websites in our survey implemented SSL, and 27\% of these sites exposed secure login forms in HTTP pages, making them vulnerable to Lupin.

\begin{table*}[!ht]
\centering
\begin{tabular}{|c|c|c|c|}
\hline
\multicolumn{2}{|c|}{Vulnerable} & \multirow{2}{*}{Not Vulnerable} & \multirow{2}{*}{Total}\\
\cline{1-2}
login form posts to HTTP& login form posts to HTTPS&  &  \\
\hline
25\% (11,313) & 3\% (1,428) & 72\% (32,255) & 100\% (45,000) \\
\hline
\end{tabular}
\caption{Distribution of websites vulnerable to Lupin}
\label{tab:crawler}
\end{table*}

\section{Defense}
\label{sec:defense}
In this section, we propose several defenses for our attack on 
password managers. First, we provide quick solutions for web 
developers to secure their login forms. Second, we propose and analyze 
several secure password manager variants; we leave it to the browser 
vendors to decide which variant is best suited for them.

\subsection{Web Application Defenses}
The most straightforward approach to defend against attacks on the 
password manager is to turn off the password manager. Websites may do 
so by setting the value of the ``autocomplete'' form attribute to  
``off''. However, this may create undesirable side effects such as 
inconveniencing users, forcing them to manually log in, as well as 
encouraging users to create less secure, easy-to-remember passwords. 

Another technique to protect HTTPS passwords from the adversary is to 
never embed a login form inside an HTTP page. If a website wishes to 
serve a portion of their content in HTTP, and switches to HTTPS for 
sensitive transactions (such as making purchases), they may do so by 
redirecting the user to a secure HTTPS login page. 

We would like to emphasize that server side solutions are not enough 
to completely mitigate the attack, since previously stored passwords 
are still vulnerable to our attack. It is essential for browser 
vendors to deploy a password manager that offers the necessary 
protections.

\subsection{Browser Defenses}

To protect users from automatic password extraction tools such as 
Lupin, Chrome and Firefox could implement defenses similar to those 
of IE and Opera. That is, they could require their password managers to be 
triggered only through user interactions. However, although this may 
mitigate the risk of automated password thefts, it is accompanied by 
usability concerns. For example, the user may now be required to 
remember the first letter of her username. Furthermore, even this 
defense does not protect against non-automated attacks. If a script is 
injected that waits patiently until a login form is filled in, then 
the attack would still succeed. In effect, this approach stops 
automated password theft attacks but not password theft attacks in 
general. 

To protect passwords submitted to web pages served in HTTPS, one could 
forbid the password manager to auto-fill any login forms containing an 
HTTPS destination address. This would consequently frustrate users 
into creating weaker passwords. Similarly, the browser could refuse to 
auto-fill passwords on HTTP pages. This would prevent a network 
attacker from obtaining any password stored by the password manager. 
However, the same disadvantages as above apply; restricting the 
password manager would only encourage users to create weaker 
passwords.

One way to achieve a balance between usability and security is use an 
approach similar to HTTP Strict Transport Security 
(HSTS)~\cite{forcehttps}. Consider an HTTP web page containing a login 
form that submits to an HTTPS page. When the user decides to store her 
password, the browser will first attempt to fetch the HTTPS version of 
the same page. If the fetch is successful, then the browser associates 
the stored password with the HTTPS version of the page. When the user 
revisits this web page, the browser will automatically redirect the 
user to the HTTPS version before auto-filling the password. One 
limitation of this defense is that it cannot protect credentials from 
pages served only in HTTP. We leave it to the browser vendors to 
decide whether this trade-off is acceptable.

\section{Related Work}
\label{sec:related}

Several researchers have attempted to design a secure password manager. However,
none of them has considered the effect of a network attacker. PwdHash transparently 
produces a different password for each site by using cryptographic hash functions~\cite{passhash}, 
hence preventing a web attacker from compromising multiple accounts from the same user
using the same password. Passpet aims to protect the user's login credentials from 
phishing attackers by associating each trusted website with user-assigned labels~\cite{passpet}. 
 
Internet users' password strength, as well as their password management habits, has
also been extensively studied in previous literature~\cite{www07_florencio,pass1,pass2,pass3}. 
Most of the existing research has found that the majority of passwords on the Internet
are weak and that users tend to reuse existing passwords. The attack described in our work does not 
target the weaknesses of these web passwords, but rather, it exploits a vulnerability in the design 
of several commercial password managers.

\section{Conclusion}
\label{sec:conclusion}
We describe an automated attack that enables a network adversary to
obtain users' credentials stored by their browsers' password managers.
To demonstrate the severity of the attack, we created a tool called 
Lupin. We evaluated Lupin in terms of its performance. For each user
on the network, Lupin is able to explore passwords stored on
1,000 websites in less than 35 seconds. In addition, we conducted
an extensive survey on the 45,000 most popular websites and discovered
that 28\% of them are vulnerable to Lupin.

{\bibliographystyle{acm}
\bibliography{paper}}

\end{document}